\documentclass[aps,apl,twocolumn,groupedaddress,amsmath]{revtex4}
\usepackage[latin1]{inputenc}    
\usepackage{graphicx}   

\begin{document}
\title{Frequency-selective single photon detection using a double quantum dot}
\author{S.~Gustavsson}
\email{simongus@phys.ethz.ch}
 \author{M.~Studer}
 \author{R.~Leturcq}
 \author{T.~Ihn}
 \author{K.~Ensslin}
 \affiliation {Solid State Physics Laboratory, ETH Z\"urich, CH-8093 Z\"urich,
 Switzerland}
\author{D. C. Driscoll}
\author{A. C. Gossard}
\affiliation{Materials Departement, University of California, Santa
Barbara, CA-93106, USA}

\date{\today}

\begin{abstract}
We use a double quantum dot as a frequency-tunable on-chip microwave
detector to investigate the radiation from electron shot-noise in a
near-by quantum point contact. The device is realized by monitoring
the inelastic tunneling of electrons between the quantum dots due to
photon absorption. The frequency of the absorbed radiation is set by
the energy separation between the dots, which is easily tuned with
gate voltages.
Using time-resolved charge detection techniques, we can directly
relate the detection of a tunneling electron to the absorption of a
single photon.
\end{abstract}

\maketitle

The interplay between quantum optics and mesoscopic physics opens up
new horizons for investigating radiation produced in nanoscale
conductors \cite{beenaaker:2001, gabelli:2004}. Microwave photons
emitted from quantum conductors are predicted to show non-classical
behavior such as anti-bunching \cite{beenakker:2004} and
entanglement \cite{emary:2005}. Experimental investigations of such
systems require sensitive, high-bandwidth detectors operating at
microwave-frequency \cite{zakka:2007}. On-chip detection schemes,
with the device and detector being strongly capacitively coupled,
offer advantages in terms of sensitivity and large bandwidths. In
previous work, the detection mechanism was implemented utilizing
photon-assisted tunneling in a
superconductor-insulator-superconductor junction \cite{deblock:2003,
onac:2006} or in a single quantum dot (QD) \cite{onac:2006b}.

\begin{figure}[htb]
\centering
 \includegraphics[width=\columnwidth]{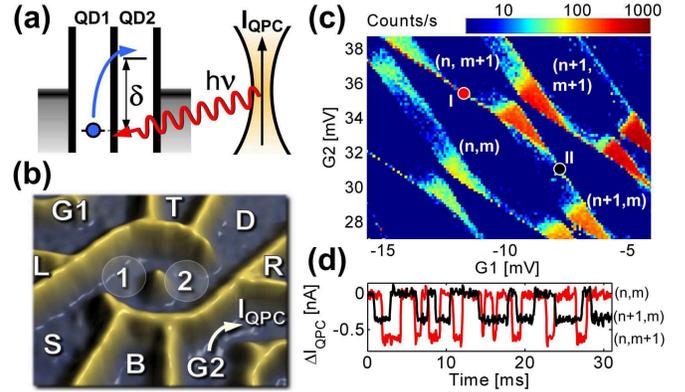}
 \caption{(a) Schematic for operating a double quantum dot (DQD) as a high-frequency noise detector. The tunable level separation
 $\delta$ of the DQD allows frequency-selective detection.
 (b) Sample used in the measurement, with two
 QDs (marked by 1 and 2) and a near-by QPC. (c)
 Charge stability diagram of the DQD, measured by counting electrons entering the
 DQD. The numbers in brackets denote the charge population of the two QDs.
 (d) Typical traces of the detector signal, taken at point I (red)
 and II (black) in (c).
 }
\label{fig:fig1}
\end{figure}

Aguado and Kouwenhoven proposed to use a double quantum dot (DQD) as
a frequency-tunable quantum noise detector \cite{aguado:2000}. The
idea is sketched in Fig.~\ref{fig:fig1}(a), showing the energy
levels of the DQD together with a quantum point contact (QPC) acting
as a noise source. The DQD is operated with a fixed detuning
$\delta$ between the electrochemical potentials of the left and
right QD.
If the system absorbs an energy $E = \delta$ from the environment,
the electron in QD1 is excited to QD2. This electron may leave to
the drain lead, a new electron enters from the source contact and
the cycle can be repeated. The process induces a current flow
through the system. Since the detuning $\delta$ may be varied
continuously by applying appropriate gate voltages, the absorbtion
energy is fully tunable. 

The scheme is experimentally challenging, due to low current levels
and fast relaxation processes between the QDs \cite{khrapai:2006}.
Here, we show that these problems can be overcome by using
time-resolved charge-detection techniques to detect single electrons
tunneling into and out of the DQD.
%
%
Apart from giving higher sensitivity than
conventional current measurement techniques, the method also allows
us to directly relate a single-electron tunneling event to the
absorbtion of a single photon. The system can thus be viewed as a
frequency-selective single-photon detector for microwave energies.
This, together with the fact that the charge-detection methods allow
precise determination of the device parameters, provide major
advantages compared to other setups \cite{gabelli:2004, zakka:2007,
deblock:2003, onac:2006, onac:2006b}.
%


The sample [Fig.~\ref{fig:fig1}(b)] was fabricated by local
oxidation \cite{fuhrer:2004} of a GaAs/Al$_{0.3}$Ga$_{0.7}$As
heterostructure, containing a two-dimensional electron gas (2DEG) 34
nm below the surface (mobility $3.5 \times 10^5~\mathrm{cm^2/Vs}$,
density $4.6\times 10^{11}~\mathrm{cm}^{-2}$). The sample also has a
backgate 1400 nm below the 2DEG, isolated by a layer of
low-temperature-grown (LT)-GaAs. The structure consists of two QDs
in series (marked by 1 and 2 in the figure) with a nearby QPC used
as a charge detector (lower-right corner of the figure). The dots
are coupled via two separate tunneling barriers, formed in the upper
and lower arms between the QDs. For this experiment, only the upper
arm was kept open, the lower one was pinched off. The gates T, B, L
and R are used to tune the height of the tunneling barriers, while
gates G1 and G2 control the electrochemical potentials of the two
QDs.

Due to electrostatic coupling between the QDs and the QPC, the
conductance of the QPC is strongly influenced by the electron
population of the QDs \cite{field:1993}. By voltage biasing the QPC
and continuously monitoring its conductance, electrons entering or
leaving the QDs can be detected in real-time \cite{vandersypen:2004,
schleser:2004, fujisawa:2004}. The time resolution is limited by the
noise of the amplifier and the capacitance of the cables, giving our
setup a bandwidth of a few kHz. Operating the QPC in a mode
analogous to the radio-frequency single electron transistor
\cite{schoelkopf:1998} should make it possible to increase the
bandwidth significantly.

The detection bandwidth puts an upper limit on the transition rates
that can be measured \cite{gustavsson:2007}. In the experiment, we
tune the tunneling rates between the QDs and the source/drain leads
to be around 1 kHz, while the coupling $t$ between the dots is kept
at a relatively large value ($t=32~\mathrm{\mu eV}$, corresponding
to $7.7~\mathrm{GHz}$). The large intradot coupling enhances the
probability for the photon absorbtion process sketched in
Fig.~\ref{fig:fig1}(a), but it also means that intradot transitions
will occur on a timescale much faster than what is detectable.

Figure \ref{fig:fig1}(c) shows a measurement of the count rate for
electrons entering the DQD versus voltages on gates $G1$ and $G2$,
with $600~\mathrm{\mu V}$ bias applied between source (S) and drain
(D). Resonant tunneling of electrons between the DQD and the source
and drain contacts give rise to lines forming a hexagon pattern. At
the crossing points of the lines, triangles with electron transport
appear due to the applied bias. These features are well-known
characteristics of DQDs and allow precise determination of the
capacitances in the system \cite{vanderwiel:2002}. The numbers in
brackets denote the charge population of the two dots. Going from
the region with population $(n,m)$ to $(n,m+1)$, resonant tunneling
occurs as QD2 aligns with the drain lead [marked by point I in Fig.
\ref{fig:fig1}(c)]. Between regions $(n,m)$ and $(n+1,m)$, the
tunneling occurs between QD1 and the source [point II]. Figure
\ref{fig:fig1}(d) displays time traces of the QPC current taken at
point I (red) and point II (black), showing a few events where
electrons enter and leave the DQD. Since the QPC is located closer
to QD2 than to QD1, electron fluctuations in QD2 give a larger
change in the QPC conductance than fluctuations in QD1. This enables
us to do charge localization measurements \cite{dicarlo:2004,
fujisawa:2006}. By analyzing the charge distribution as a function
of detuning $\delta$, we extract the tunnel coupling energy between
the QDs to be $t = 32~\mathrm{\mu eV}$ \cite{dicarlo:2004}.


%

\begin{figure}[tb]
\centering
 \includegraphics[width=\columnwidth]{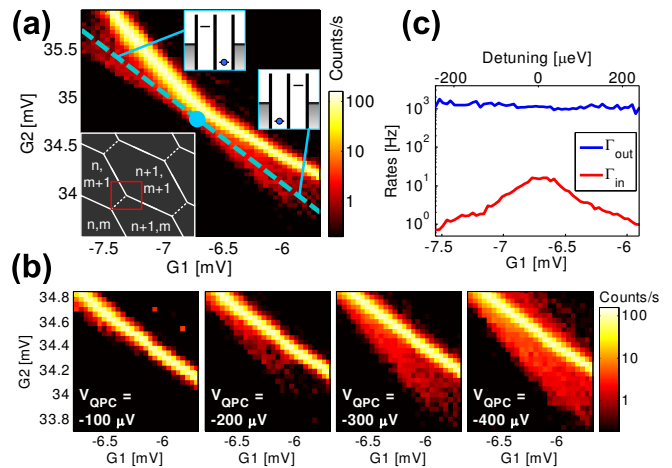}
 \caption{(a) Electron count rates for a small region close to
 a triple point (marked by a blue point). The inset shows a sketch of the surrounding hexagon pattern.
 The dashed line denotes the detuning axis, with zero detuning occurring at the triple
 point. The data was taken with $V_{QPC}=-300~\mathrm{\mu V}$.
 (b) Blow-up of the lower-right region of (a), measured for
 different QPC bias voltages.
 (c) Rates for electron tunneling into and out of the DQD, measured
 along the dashed line in (a). $\Gamma_{\mathrm{in}}$ falls of rapidly with detuning, while $\Gamma_{\mathrm{out}}$
 shows only minor variations.
   }
\label{fig:fig2}
\end{figure}

In the following, we present measurements taken with zero bias
across the DQD. Fig.~\ref{fig:fig2}(a) shows count rates close to
the triple point where the $(n+1,m)$, $(n,m+1)$ and $(n+1,m+1)$
states are degenerate [see inset of Fig.~\ref{fig:fig2}(a)]. The
arguments presented below are applicable also for the triple point
between the $(n,m)$, $(n+1,m)$, $(n,m+1)$ states, but for simplicity
we consider only the first case. At the triple point [marked by a
blue dot in Fig.~\ref{fig:fig2}(a)], the detuning $\delta$ is zero
and both dots are aligned with the Fermi level of the leads.
The two strong, bright lines emerging from this point come from
resonant tunneling between the left (right) QD and the source
(drain) lead. The height of the lines gives directly the strength of
the tunnel couplings \cite{schleser:2004, naaman:2006}, and we find
the rates to be $\Gamma_\mathrm{S} = 1.2~\mathrm{kHz}$ and
$\Gamma_\mathrm{D} = 1.1~\mathrm{kHz}$.

Along the blue dashed line in Fig.~\ref{fig:fig2}(a), there are
triangle-shaped regions with low but non-zero count rates where
tunneling is expected to be strongly suppressed due to Coulomb
blockade. The DQD level arrangement inside the triangles is shown in
the insets. Comparing with the sketch in Fig.~\ref{fig:fig1}(a), we
see that both regions have DQD configurations favorable for noise
detection. The dashed blue line connecting the triangles defines the
detuning axis, with zero detuning occuring at the triple point. We
take detuning to be negative in the upper-left part of the figure.
In Fig.~\ref{fig:fig2}(b), the lower-right part of
Fig.~\ref{fig:fig2}(a) was measured for four different QPC bias
voltages. The resonant line stays the same in all four measurements,
but the triangle becomes both larger and more prominent as the QPC
bias is increased. This is a strong indication that the tunneling is
due to absorbtion of energy from the QPC.

\begin{figure}[tb]
\centering
 \includegraphics[width=\columnwidth]{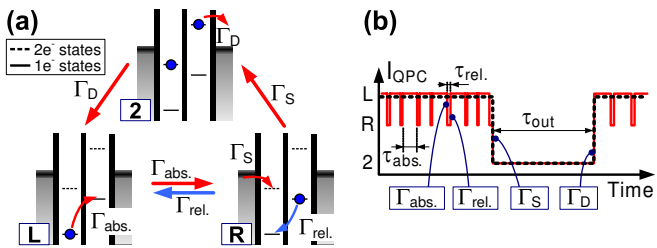}
 \caption{(a) Energy level diagrams for the three states of the DQD. The labels $L$, $R$ and $2$
 denote the excess charge population. The levels are raised by the intradot charging energy
 $E_{Ci}$ when the DQD holds two excess electrons.
 (b) Schematic changes of the detector signal as electrons tunnel into, between and out of the DQD.
 }
\label{fig:fig3}
\end{figure}

The time-resolved measurement technique allows the rates for
electron tunneling into and out of the DQD to be determined
separately \cite{gustavsson:2005}. Figure \ref{fig:fig2}(c) shows
the rates $\Gamma_{\mathrm{in}}$ and $\Gamma_{\mathrm{out}}$
measured along the dashed line of Fig.~\ref{fig:fig2}(a). The rate
for tunneling out stays almost constant along the line, but
$\Gamma_{\mathrm{in}}$ is maximum close to the triple point and
falls of rapidly with increased detuning. This suggests that only
the rate for electrons tunneling into the DQD is related to the
absorbtion process. To explain the experimental findings we model
the system using a rate-equation approach. For a configuration
around the triple point, the DQD may hold $(n+1,m)$, $(n,m+1)$ or
$(n+1,m+1)$ electrons. We label the states $L$, $R$ and $2$ and draw
the energy diagrams together with possible transitions in
Fig.~\ref{fig:fig3}(a). The figure shows the case for positive
detuning, with $\delta \gg k_B T$. Note that when the DQD holds two
excess electrons, the energy levels are raised by the intradot
charging energy, $E_{Ci}=800~\mathrm{\mu eV}$.

In Fig \ref{fig:fig3}(b) we sketch the time evolution of the system.
The red curve shows the expected charge detector signal assuming a
detector bandwidth much larger than the transitions rates. Starting
in state $L$, the electron is trapped until it absorbs a photon and
is excited to state $R$ (with rate $\Gamma_{\mathrm{abs.}}$). From
here, the electron may either relax back to state $L$ (rate
$\Gamma_{\mathrm{rel.}}$) or a new electron may enter QD1 from the
source lead and put the system into state $2$ (rate
$\Gamma_{\mathrm{S}}$). Finally, if the DQD ends up in state $2$,
the only possible transition is for the electron in the right dot to
leave to the drain lead.

The relaxation rate for a similar DQD system has been measured to be
$1/\Gamma_{\mathrm{rel.}} = 16~\mathrm{ns}$ \cite{petta:2004}, which
is much faster than the available measurement bandwidth. Therefore,
the detector will not be able to register the transitions where the
electron is repeatedly excited and relaxed between the dots. Only
when a second electron enters from the source lead [transition
marked by $\Gamma_{\mathrm{S}}$ in Fig.~\ref{fig:fig3}(a, b)], the
DQD will be trapped in state $2$ for a sufficiently long time
($\sim\! 1/\Gamma_D \sim\! 1~\mathrm{ms}$) to allow detection. The
measured time trace will only show two levels, as indicated by the
dashed line in Fig.~\ref{fig:fig3}(b). Such a trace still allows
extraction of the effective rates for electrons entering and leaving
the DQD, $\Gamma_{\mathrm{in}} = 1/\langle \tau_{\mathrm{in}}
\rangle$ and $\Gamma_{\mathrm{out}} = 1/\langle \tau_{\mathrm{out}}
\rangle$. To relate $\Gamma_{\mathrm{in}}$, $\Gamma_{\mathrm{out}}$
to the internal DQD transitions, we write down the Master equation
for the occupation probabilities of the states:
\begin{equation}\label{eq:master} \frac{d}{dt }\left(
\begin{array}{c}
             p_L \\
             p_R \\
              p_2 \\
              \end{array} \right) =
                        \left(
          \begin{array}{ccc}
            -\Gamma_{\mathrm{abs.}} & \Gamma_{\mathrm{rel.}} & \Gamma_{\mathrm{D}} \\
            \Gamma_{\mathrm{abs.}} & -(\Gamma_{\mathrm{S}}+\Gamma_{\mathrm{rel.}}) & 0 \\
            0 & \Gamma_{\mathrm{S}} & -\Gamma_{\mathrm{D}} \\
          \end{array}
        \right) \left(
                        \begin{array}{c}
                          p_L \\
                          p_R \\
                          p_2 \\
                        \end{array}
                      \right).
\end{equation}
Again, we assume positive detuning, with $\delta \gg k_B T$. The
measured rates $\Gamma_{\mathrm{in}}$, $\Gamma_{\mathrm{out}}$ are
calculated from the steady-state solution of Eq.~\ref{eq:master}:
\begin{eqnarray}
  \Gamma_{\mathrm{in}} &=& \Gamma_{\mathrm{S}} \, \frac{p_R}{p_L+p_R} =
  \frac{\Gamma_{\mathrm{S}} \Gamma_{\mathrm{abs.}}}
  {\Gamma_{\mathrm{S}} + \Gamma_{\mathrm{abs.}} + \Gamma_{\mathrm{rel.}}}, \\
  \Gamma_{\mathrm{out}} &=& \Gamma_{\mathrm{D}}. \label{eq:GoutEqGsGd}
\end{eqnarray}
In the limit $\Gamma_{\mathrm{rel.}} \gg \Gamma_{\mathrm{S}},\,
\Gamma_{\mathrm{abs.}}$, the first expression simplifies to
\begin{equation}\label{eq:GinGabs}
 \Gamma_{\mathrm{in}}=\Gamma_{\mathrm{S}} \,
 \Gamma_{\mathrm{abs.}}/\Gamma_{\mathrm{rel.}}.
\end{equation}
The corresponding expressions for negative detuning are found by
interchanging $\Gamma_{\mathrm{S}}$ and $\Gamma_{\mathrm{D}}$ in
Eqs.~(2-4). Coming back to the experimental findings of
Fig.~\ref{fig:fig2}(c), we note that $\Gamma_\mathrm{out}$ only
shows small variations within the region of interest. This together
with the result of Eq.~(\ref{eq:GoutEqGsGd}) suggest that we can
take $\Gamma_\mathrm{S}$, $\Gamma_\mathrm{D}$ to be independent of
detuning.
The rate $\Gamma_\mathrm{in}$ in Eq.~(\ref{eq:GinGabs}) thus
reflects the dependence of
$\Gamma_\mathrm{abs.}/\Gamma_\mathrm{rel.}$ on detuning. Assuming
also $\Gamma_\mathrm{rel.}$ to be constant, a measurement of
$\Gamma_\mathrm{in}$ gives directly the absorbtion spectrum of the
DQD. The measurements cannot exclude that $\Gamma_\mathrm{rel.}$
also varies with $\delta$, but as we show below the model assuming
$\Gamma_\mathrm{rel.}$ independent of detuning fits the data well.

Equation~(\ref{eq:GinGabs}) shows that the low-bandwidth detector
can be used to measure the absorbtion spectrum, even in the presence
of fast relaxation. Moreover, the detection of an electron entering
the DQD implies that a quantum of energy was absorbed immediately
before the electron was detected. The charge detector signal thus
relates directly to the detection of a single photon.

\begin{figure}[htb]
\centering
 \includegraphics[width=\columnwidth]{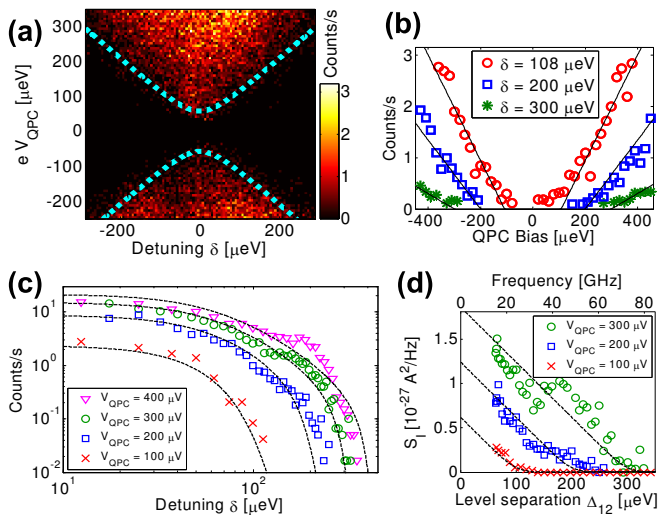}
 \caption{Count rate measured versus detuning and QPC bias voltage.
 The dashed line shows the level separation for a two-level
 system, with $\Delta_{12}= \sqrt{4\,t^2 + \delta^2}$.
 There are only counts in the region where $|e V_{QPC}| > \Delta_{12}$.
 (b) Count rate versus QPC bias for different values of
 detuning. The solid lines are guides to the eye.
 (c) DQD absorption spectrum, measured for different QPC bias. The
  dashed lines are the results of Eq.~(\ref{eq:absRate}), with
 parameters given in the text.
 (d) Noise spectrum of the QPC, extracted from the data in (c). The
 dashed lines show spectra expected from Eq.~(\ref{eq:SI}).
 }
\label{fig:fig4}
\end{figure}

In the following, we use the DQD to quantitatively investigate the
microwave radiation emitted from the nearby QPC. Figure
\ref{fig:fig4}(a) shows the measured $\Gamma_{\mathrm{in}}$ versus
detuning and QPC bias. The data was taken along the dashed line of
Fig.~\ref{fig:fig2}(a), with gate voltages converted into energy
using lever arms extracted from finite bias measurements. Due to the
tunnel coupling $t$ between the QDs, the energy level separation
$\Delta_{12}$ of the DQD is given by $\Delta_{12} =\sqrt{4\,t^2 +
\delta^2}$. The dashed lines in \ref{fig:fig4}(a) show
$\Delta_{12}$, with $t = 32~\mathrm{\mu eV}$. A striking feature is
that there are no counts in regions with $|eV_{QPC}| < \Delta_{12}$.
This originates from the fact that the voltage-biased QPC can only
emit photons with energy $\hbar \omega \le eV_{QPC}$
\cite{aguado:2000, onac:2006b, zakka:2007}. The result presents
another strong evidence that the absorbed photons originate from the
QPC.


To describe the results quantitatively, we consider the emission
spectrum of a voltage biased QPC with one conducting channel. In the
low-temperature limit $k_B T \ll \hbar \omega  $, the spectral noise
density $S_I(\omega)$ for the emission side ($\omega>0$) takes the
form (see \cite{aguado:2000} for the full expression)
\begin{equation}\label{eq:SI}
 S_I(\omega) =  \frac{4 e^2}{h} D (1-D) \frac{e V_{QPC} - \hbar \omega}{1-e^{-(e V_{QPC} -  \hbar \omega)/k_B
 T}},
\end{equation}
where $D$ is the transmission coefficient of the channel.
Using the model of Ref. \cite{aguado:2000}, we find the absorption
rate of the DQD in the presence of the QPC:
\begin{equation}\label{eq:absRate}
 \Gamma_\mathrm{abs.} = \frac{4 \pi e^2 k^2 t^2 Z_l^2}{h^2}
 \frac{S_I(\Delta_{12}/\hbar)}{\Delta_{12}^2}.
\end{equation}
The constant $k$ is the capacitive lever arm of the QPC on the DQD
and $Z_l$ is the zero-frequency impedance of the leads connecting
the QPC to the voltage source.
Equation (\ref{eq:absRate}) states how well fluctuations in the QPC
couple to the DQD system.

Figure \ref{fig:fig4}(b) shows the measured absorbtion rates versus
$V_{QPC}$, taken for three different values of $\delta$. As expected
from Eqs.~(\ref{eq:SI}, \ref{eq:absRate}), the absorption rates
increase linearly with bias voltage as soon as $|eV_{QPC}| >
\delta$. The different slopes for the three data sets are due to the
$1/\Delta_{12}^2$-dependence in the relation between the emission
spectrum and the absorption rate of Eq.~(\ref{eq:absRate}). In
Fig.~\ref{fig:fig4}(c), we present measurements of the absorption
spectrum for fixed $V_{QPC}$. The rates decrease with increased
detuning, with sharp cut-offs as $|\delta| > e V_{QPC}$. In the
region of small detuning, the absorption rates saturate as the DQD
level separation $\Delta_{12}$ approaches the limit set by the
tunnel coupling. The dashed lines show the combined results of
Eqs.~(\ref{eq:GinGabs}-\ref{eq:absRate}), with parameters
$T=0.1~\mathrm{K}$, $Z_l = 0.7~\mathrm{k\Omega}$, $D=0.5$,
$t=32~\mathrm{\mu eV}$, $k = 0.15$, $\Gamma_{\mathrm{S}} =
1.2~\mathrm{kHz}$ and $\Gamma_{\mathrm{D}} = 1.1~\mathrm{kHz}$.
Using $\Gamma_{\mathrm{rel.}}$ as a fitting parameter, we find
$1/\Gamma_{\mathrm{rel.}} = 5~\mathrm{ns}$. This should be seen as a
rough estimate of $\Gamma_\mathrm{rel.}$ due to uncertainties in
$Z_l$, but it shows reasonable agreement with previously reported
measurements \cite{petta:2004}. The overall good agreement between
the data and the electrostatic model of Eq.~(\ref{eq:absRate})
supports the assumption that the interchange of energy between the
QPC and the DQD is predominantly mediated by photons instead of
phonons or plasmons.

The data for $V_{QPC}=400~\mu V$ shows some irregularities compared
to theory, especially at large positive detuning. We speculate that
the deviations are due to excited states of the individual QDs, with
excitation energies smaller than the detuning. In
Fig.~\ref{fig:fig4}(d), we convert the detuning $\delta$ to level
separation $\Delta_{12}$ and use Eq.~(\ref{eq:absRate}) to extract
the noise spectrum $S_I$ of the QPC. The linear dependence of the
noise with respect to frequency corresponds well to the behavior
expected from Eq.~(\ref{eq:SI}). Again, the deviations at
$\Delta_{12}=190~\mathrm{\mu eV}$ are probably due to an excited
state in one of the QDs. The single-level spacing of the QD is
$\Delta E \approx 200~\mathrm{\mu eV}$, which sets an upper bound on
frequencies that can be detected with this method. The
frequency-range can be extended by using DQD in carbon nanotubes
\cite{mason:2004} or InAs nanowires \cite{fasth:2005, pfund:2006},
where the single-level spacing is significantly larger.

To summarize, we have shown that a DQD can be used as a
frequency-selective detector for microwave radiation. Time-resolved
charge detection techniques allow single photons to be detected,
giving the method a very high sensitivity. The ability to detect
single photons also opens up the possibility to investigate the
statistics of the absorbed radiation. By fabricating a pair of DQD
devices and investigating the cross-correlations, time-dependent
photon correlations can be directly measured \cite{beenaaker:2001}.
To prove the principle of the device we have investigated the
high-frequency spectrum of radiation emitted from a voltage-biased
QPC. The emission rate was found to increase linearly with applied
bias, with a spectrum having a sharp cut-off for frequencies higher
than the QPC bias.



\bibliographystyle{apsrev}
\bibliography{PhotonDetection}

\end{document}